\documentclass[aps,prd,reprint,preprintnumbers,superscriptaddress,showpacs,twocolumn]{revtex4-1}
\usepackage{latexsym,graphicx,amssymb,amsmath,mathrsfs}
\usepackage{setspace,bm}
\usepackage[breaklinks, colorlinks=true, pdfstartview=FitV, linkcolor=red, citecolor=blue, urlcolor=blue]{hyperref}
\usepackage{amsmath}
\usepackage[usenames]{color}
\usepackage{latexsym}
\usepackage{epstopdf}
\usepackage{amssymb}

\newcommand{\Slash}[1]{{\ooalign{\hfil/\hfil\crcr$#1$}}}
\newcommand\nc{N_\mathrm{c}}
\newcommand\nf{N_\mathrm{f}}

\usepackage[normalem]{ulem}  

\newcommand{\comment}[1]{}

\renewcommand\sout{\bgroup \color{red} \ULdepth=-.5ex \ULset}

\begin{document}
\preprint{YITP-19-10, KUNS-2749}

\title{Application of the path optimization method to the sign problem
\\ in an effective model of QCD with a repulsive vector-type interaction}

\author{Kouji Kashiwa}
\email[]{kashiwa@fit.ac.jp}
\affiliation{Fukuoka Institute of Technology, Wajiro, Fukuoka 811-0295,
Japan}

\author{Yuto Mori}
\email[]{mori@ruby.scphys.kyoto-u.ac.jp}
\affiliation{Department of Physics, Faculty of Science, Kyoto
University, Kyoto 606-8502, Japan}

\author{Akira Ohnishi}
\email[]{ohnishi@yukawa.kyoto-u.ac.jp}
\affiliation{Yukawa Institute for Theoretical Physics,
Kyoto University, Kyoto 606-8502, Japan}

\begin{abstract}
The path optimization method is applied to a QCD effective model with the
 Polyakov loop and the repulsive vector-type interaction at finite
 temperature and density to circumvent the model sign problem.
We show how the path optimization method can increase the average phase
 factor and control the model sign problem.
This is the first study which correctly treats the repulsive
 vector-type interaction in the QCD effective model with the Polyakov-loop
 via the Markov-chain Monte-Carlo approach.
It is shown that we can evade the model sign problem 
within the standard path-integral formulation
by complexifying the temporal component of the gluon field
and the vector-type auxiliary field.
\end{abstract}

\maketitle

\section{Introduction}
Understanding the confinement-deconfinement transition at finite
temperature ($T$) and chemical potential ($\mu$) in quantum
chromodynamics (QCD) is one of important and interesting subjects
in elementary particle, nuclear and hadron physics.
To investigate non-perturbative properties of QCD such as
the chiral symmetry breaking and the confinement mechanism,
Monte-Carlo simulations of lattice QCD
have been utilized as a powerful tool
for studying nonperturbative properties of QCD
such as the chiral symmetry breaking and the confinement
at zero baryon density.
Unfortunately, lattice QCD simulations have the sign problem 
at nonzero real quark chemical potential.
To circumvent the sign problem, several methods
have been proposed so far such as the Taylor expansion
method~\cite{Miyamura:2002en,Allton:2005gk,Gavai:2008zr},
the reweighting
method~\cite{Fodor:2001au,*Fodor:2001pe,*Fodor:2004nz,Fodor:2002km},
the analytic continuation
method~\cite{deForcrand:2002ci,*deForcrand:2003hx,D'Elia:2002gd,*D'Elia:2004at,Chen:2004tb},
and the canonical
approach~\cite{Hasenfratz:1991ax,Alexandru:2005ix,Kratochvila:2006jx,deForcrand:2006ec,Li:2010qf,Nakamura:2015jra}.
However, we cannot access cold dense region, $\mu/T > 1$, in these methods
at present~\cite{deForcrand:2010ys}.

The QCD effective models are widely used to investigate the QCD phase structure
at finite real chemical potential.
We can sometimes avoid the sign problem
in simple models such as
the Nambu--Jona-Lasinio (NJL) model
without the repulsive vector-type interaction.
In more realistic models, however,
the sign problem arises again.
For example, the Polyakov-loop extended NJL (PNJL)
model~\cite{Fukushima:2003fw}
has the sign problem even in the mean-field treatment.
The sign problem appearing in the QCD effective model is
called the model sign problem~\cite{Fukushima:2006uv,Tanizaki:2015pua}.
Practically, one can avoid the model sign problem
by using some prescriptions,
which may not have clear theoretical foundation.

One of the model sign problems arises from the vector-type interaction.
In the mean-field approximation for the NJL-type model,
we usually neglect the Wick rotation of
the vector-type auxiliary-field ($\omega_0$)
and then the stationary point of the action is
considered to be the solution.
The stationary point is corresponding
to the maxima of the thermodynamic potential
along the $\omega_0$ direction and thus it is not stable in principle.
While this treatment cannot be justified from the standard path-integral
formulation,
it can be acceptable in the mean-field approximation.
Actually, this problem has been discussed in Ref.~\cite{Mori:2017zyl}
by using the Lefschetz thimble
method~\cite{Witten:2010cx,Cristoforetti:2012su,Fujii:2013sra}.
We can clearly understand that the standard mean-field
treatment implicitly employs the complexification of the vector-type
auxiliary-field based on the Cauchy(-Poincare) theorem.

In this article, we use the path optimization
method~\cite{Mori:2017pne,Ohnishi:2017zxh,Mori:2017nwj,Kashiwa:2018vxr}
to formally tackle
the model sign problem induced by the Polyakov-loop and also the
repulsive vector-type interaction in a QCD effective model.
In Ref.~\cite{Kashiwa:2018vxr}, we have shown that the
complexification of the temporal component of the gluon field is
sufficient to control the model sign problem in the PNJL model
without the repulsive vector-type interaction.
In addition, we have shown that the complexification of
the vector-type auxiliary field should be responsible
to control the model sign problem in the NJL model
with the repulsive vector-type interaction~\cite{Mori:2017zyl}.
It should be noted that the flow equation of the Lefschetz thimble
blows up at a small value of the vector-type auxiliary field,
and we failed to obtain the Lefschetz thimbles
in the auxiliary-field space in Ref.~\cite{Mori:2017zyl}.
Therefore, in this article, we apply the path optimization method to the PNJL model with
the repulsive vector-type interaction to control
its model sign problem.
This study is the first attempt to treat the model sign problem
correctly and systematically
within the standard path-integral formulation via
the complexification of the integral variables in the QCD effective
model with the Polyakov loop and the repulsive vector-type interaction.

This article is organized as follows.
In the next section, we explain the path optimization method and the
PNJL model with the repulsive vector-type interaction.
Section~\ref{Sec:NR} shows numerical results by using the hybrid
Monte-Carlo method.
Section~\ref{Sec:Summary} is devoted to summary and discussions.

\section{Formulation}
\label{Sec:F}

We investigate the model sign problem appearing in
the PNJL model with the repulsive vector-type interaction via the path
optimization method.
Details are explained below.

 \subsection{Polyakov-loop extended Nambu--Jona Lasinio model}

The Euclidean Lagrangian density of the two-flavor PNJL
model~\cite{Fukushima:2003fw} with
the repulsive vector-type interaction is given as
\begin{align}
 {\cal L}
 &= {\bar q} (\Slash{D} + m_0)q
  - G[({\bar q}q)^2+({\bar q} i\gamma_5 {\vec \tau} q)^2]
 + G_v ({\bar q} \gamma_\mu q)^2
 \nonumber\\
 &+ {\cal V}_{\mathrm{g}} (\Phi,{\bar \Phi}),
\end{align}
where $m_0$ denotes the current quark mass,
$D_\nu= \partial_\nu - i g A_\nu \delta_{\nu 4}$ is the covariant derivative,
$\Phi$ (${\bar \Phi}$) represents the Polyakov-loop (its conjugate),
and ${\cal V}_{\mathrm{g}}$ expresses the gluonic contribution.
The coupling constants $G$ and $G_v$ take positive values,
as understood from the QCD one-gluon exchange
interaction, see Ref.~\cite{Kashiwa:2011td} as an example.

We employ the homogeneous auxiliary-field ansatz,
as adopted
in the previous works
using the Monte-Carlo PNJL model~\cite{Cristoforetti:2010sn,Kashiwa:2018vxr},
and thus our numerical results converge to the mean-field results in the
infinite volume limit.
The homogeneous ansatz corresponds to the momentum truncation
to $\bold{k}=\bold{0}$.
After the bosonization and complexification of auxiliary fields,
the grand-canonical partition function is given as
\begin{align}
{\cal Z} &= \int \prod_{{\bf k}} d z_{\bf k}~e^{-\Gamma[z_{\bf k}]},
\end{align}
where
$\Gamma$ is the effective action,
and $z_\mathbf{k}$ represents the dynamical variables in the momentum space.
With the homogeneous field ansatz, we truncate the auxiliary fields to
${\bf k}=0$ components.
Then the effective action becomes
$\Gamma=\beta V {\cal V}$ where ${\cal V}$ is the effective potential
and $\beta$ is the inverse temperature.
Thus our Monte-Carlo results should agree with the mean-field results in
the infinite volume limit, where the configuration
at the minimum of $\mathcal{V}$ dominates.

After the  Hubbard-Stratonovich transformation (bosonization),
the thermodynamic potential is obtained as
\begin{align}
 {\cal V} &= {\cal V}_{\mathrm{NJL}} + {\cal V}_{\mathrm{g}},
\end{align}
where ${\cal V}_\mathrm{\mathrm{NJL}}$ and ${\cal V}_\mathrm{g}$
are the fermionic and gluonic
parts of the effective potential, respectively.
The actual form of ${\cal V}_\mathrm{NJL}$ is given as
\begin{align}
 {\cal V}_\mathrm{NJL}
 &= - 2 \nf \int_\Lambda \frac{d^3 p}{(2\pi)^3}
      \Bigl[ \nc E_{\bf p} - \nc \sqrt{{\bf p}^2 + m_0^2}
\nonumber\\
         &+ T \ln \Bigl( f^- f^+ \Bigr)\Bigr]
 + G(\sigma^2 + {\vec \pi}^2) + G_v \omega_4^2,
 \label{Eq:NJL}
\end{align}
where $\nf=2$ ($\nc=3$) is the number of flavor (color)
and $\Lambda$ is the three-dimensional momentum cutoff.
We set the same momentum cutoff in the vacuum and the thermal parts.
We here introduce auxiliary fields as $\sigma={\bar q} q$,
${\vec \pi} = {\bar q} i\gamma_5 \vec{\tau}q$ and
$\omega_4 = - {\bar q} i\gamma_0 q$.
The Fermi-Dirac distribution functions are given as
\begin{align}
&f^-
= 1
 + 3 (\Phi+{\bar \Phi} e^{-\beta E_{\bf p}^-} ) e^{-\beta E_{\bf p}^-}
 + e^{-3\beta E_{\bf p}^-},
\nonumber\\
&f^+
= 1
 + 3 ({\bar \Phi}+\Phi e^{-\beta E_{\bf p}^+} ) e^{-\beta E_{\bf p}^+}
 + e^{-3\beta E_{\bf p}^+},
\label{Eq:FD}
\\
&E_{\bf p}^\mp = E_{\bf p} \mp {\tilde \mu}
                    = \sqrt{\varepsilon_{\bf p}^2 + 2 N^+N^-} \mp
                    {\tilde \mu}
,\nonumber\\
&\varepsilon_{\bf p} = \sqrt{{\bf p}^2 + M^2+ N^2},
\nonumber
\end{align}
where $M, N, N^\pm$ and $\tilde{\mu}$ are functions of the
auxiliary fields,
\begin{align}
 &M = m_0 - 2 G \sigma,~~~N = - 2 G \pi^0,~~~N^\pm = - 2 G \pi^\pm,
 \nonumber\\
 &{\tilde \mu} = \mu - 2 i G_v \omega_4,
\end{align}
with $\pi^0 = \pi_3$ and $\pi^\pm = (\pi_1 \pm i \pi_2)/ \sqrt{2}$.
Because of the $ 2 i G_v \omega_4$ term in ${\tilde \mu}$,
the repulsive vector-type interaction
induces the model sign problem in addition to that from the Polyakov-loop.
For $\mathcal{V}_\mathrm{g}$,
we employ the logarithmic type Polyakov-loop potential
proposed in Ref.~\cite{Roessner:2006xn},
\begin{align}
&\frac{{\cal V}_{g}}{T^4}
= - \frac{1}{2} a_T {\bar \Phi} \Phi + b_T \ln (h),
 \label{Eq:PP}
\\
&h = 1 - 6 {\bar \Phi} \Phi
       + 4 ({\bar \Phi}^3 + \Phi^3)
       - 3 ({\bar \Phi} \Phi)^2,
\\
& a_T = a_0 + a_1 \Bigl(\frac{T_0}{T} \Bigr)
            + a_2 \Bigl(\frac{T_0}{T} \Bigr)^2,~~~
 b_T  = b_3 \Bigl(\frac{T_0}{T} \Bigr)^3.
\end{align}
The parameters are usually set to reproduce the lattice QCD data in the pure
gauge limit.
Basic setup to compute the PNJL model with the Markov-Chain Monte-Carlo
method is shown in Refs.~\cite{Cristoforetti:2010sn,Kashiwa:2018vxr}.

Cuts in the logarithm of Eq.~(\ref{Eq:PP}) may induce the numerical
problem but it may be the model artifact and thus we do not consider any
additional care for the singularities in this study as in
Ref.~\cite{Kashiwa:2018vxr}.
One of the promising ways to avoid the problem is the modification
of the functional form of the Polyakov-loop potential.
The logarithmic term in the Polyakov-loop potential appears as
$ V T_0^3 b_3 \times \ln (h) $ in the Boltzmann weight.
If $VT_0^3 b_3$ is set to be a positive integer,
the singularity does not matter.
In the present potential, $V T_0^3 b_3$ is not an integer.
It is well known that there is another functional form of the Polyakov-loop potential
that is the polynomial one~\cite{Ratti:2005jh}, which also reproduces
the lattice QCD data in the pure gauge limit at finite $T$
and
does not have singularities.
Nevertheless,
sampled configurations are found to be well localized, then
the path optimization method
works well practically in the present setup as shown later.

 \subsection{Path optimization method}

In the path optimization method, we first
complexify the integral variables,
$x_i \in \mathbb{R} \to z_i \in \mathbb{C}$
where $i=1, \cdots, n$ with $n$ being the dimension of integration.
To construct the new (and good)
integral path in the complex space, we use
the cost function which represents the seriousness of the sign problem.
We vary the integral path in the direction
to decrease the cost function.
This method has similarity from the viewpoint of the
complexification of dynamical variables with the complex Langevin
method~\cite{Parisi:1980ys,Parisi:1984cs} and the
Lefschetz thimble method~\cite{Witten:2010cx,Cristoforetti:2012su,Fujii:2013sra,Alexandru:2015sua}.
Especially, the path optimization method belongs to the category
of the off-thimble integral methods,
which allow the integral path to deviate from the thimble,
as proposed in the generalized Lefschetz-thimble
method~\cite{Alexandru:2015sua}.
See
Refs.~\cite{Nishimura:2017eiu,Ito:2016efb,DiRenzo:2016pwd,DiRenzo:2017igr,Attanasio:2018rtq,Bursa:2018ykf,Alexandru:2018fqp}
for recent progress in these methods.

The path optimization method was firstly proposed in Ref.\,\cite{Mori:2017pne}.
The machine learning (feedforward neural-network)
was introduced to describe and to optimize the modified integral path
in Refs.\,\cite{Mori:2017nwj,Ohnishi:2017zxh}.
Few days before Ref.~\cite{Mori:2017nwj} was submitted,
the machine learning was introduced to learn the integral manifold in
the generalized Lefschetz-thimble method in Ref.\,\cite{Alexandru:2017czx}.
This method uses
the supervised learning because we must teach the relevant integral path
(manifold) to the neural-network, and the results of the generalized
Lefschetz thimble method have been used as the teacher data;
it is the first paper which employs
the supervised learning to evade the sign problem as far as we know.
Also, the same group applied the machine learning to optimize the integral
path by using the average phase factor
in Ref.\,\cite{Alexandru:2018fqp} after our paper \cite{Mori:2017nwj} appeared.
This method has similarity with our path optimization method
which uses the unsupervised learning.
The machine learning can be applied to various
optimization problems and thus it is quite useful in physics.

The functional form of the new integral path is represented by
using the feedforward neural network~\cite{Mori:2017pne,Mori:2017nwj}.
Then, the parameters in the feedforward neural network are
optimized via the minimization of the cost function.
The largest advantage of using the feedforward neural network
in the path optimization method is
in the universal approximation theorem;
the neural network even with the mono hidden-layer can
approximate any kind of continuous function on the compact subset as
long as we
prepare sufficient number of units in the
hidden layer~\cite{cybenko1989approximation,hornik1991approximation}.

To use the feedforward neural network, we represent $z_i$ by using
parametric quantity ($t_i$) as
 \begin{align}
 a_i(t) &= g(w^{(1)}_{ij} t_j + b^{(1)}_i),~~~~
 f_i(t)  = g(w^{(2)}_{ij} a_j + b^{(2)}_i),
 \nonumber\\
 z_i(t) &= t_i + i [ \alpha_i f_i(t) + \beta_i ],
 \label{Eq:fo}
 \end{align}
where $w_{ij}$, $b_i$, $\alpha_i$ and $\beta_i$ are parameters.
In particular, $w$ and $b$ are so called the weight and the bias,
respectively.
Thus, we have the map $\mathrm{Re}(z_i) \twoheadrightarrow
\mathrm{Im}(z_i)$.
The function $g(x)$ is  so called the activation function and we
use the hyperbolic tangent function.
We use the back-propagation algorithm in actual optimization of
parameters.
It should be noted that the path optimization method reproduces the same
results with the original theory because of the Cauchy(-Poincare)
theorem as long as the integral path does not go across singular points
and the contribution at $\mathrm{Re}~z \to \pm \infty$ vanishes.

To obtain the good integral path, we use the following form of the
cost function;
\begin{align}
 {\cal F}[z(t)]
 &= \frac{1}{2} \int d^nt~
   |e^{i \theta(t)} - e^{i \theta _0}|^2 \times |J(t) e^{-\Gamma(z(t))}| \nonumber \\
 &= \int d^nt~|J(t) e^{-\Gamma(z(t))}|
  \nonumber\\
 & \hspace{1.5cm}
    - \left |\int d^nt~J(t) e^{-\Gamma(z(t))} \right |,
\label{Eq:cf}
\end{align}
where
\begin{align}
 \theta(t) &= \arg (J(t)e^{-\Gamma(z(t))}), ~~~
 \theta_0   = \arg ({\cal Z}),
 \nonumber\\
 J(t) &= \det \Bigl( \frac{\partial z_i}{\partial t_j} \Bigr),
\end{align}
see Ref.~\cite{Mori:2017pne} for
details.

In applying the path optimization method to the PNJL model
with the vector-type interaction,
we complexify temporal gluon components ($A_3$ and $A_8$)
and the Wick rotated vector-type auxiliary field ($\omega_4$),
while
the scalar-type and pseudo-scalar-type auxiliary
fields ($\sigma$ and ${\vec \pi}$) are still treated as real variables.
Thus, we have
7 dynamical variables
($\sigma$, $\pi^0$, $\pi^\pm$, $\mathrm{Re}\, \omega_4$, $\mathrm{Re}\,
A_3$, $\mathrm{Re}\, A_8$) and
3 dependent variables ($\mathrm{Im}\, \omega_4$, $\mathrm{Im}\, A_3$,
$\mathrm{Im}~A_8$) where
latter three variables are given via Eq.~(\ref{Eq:fo}).
Since it is known that the model sign problem can be resolved by the
complexification of the temporal gluon fields in the Lefschetz-thimble
method at least in the system without the
repulsive vector-type interaction~\cite{Tanizaki:2015pua}.
Also, $\omega_4$ can induce the model sign problem even in the NJL model
and then we must consider the complexification of
$\omega_4$~\cite{Mori:2017zyl}.

In the present study, we directly complexify $A_3$, $A_8$ and
$\omega_4$, but this treatment may violate
the periodicity along the
$\mathrm{Re}\, A_3$ and $\mathrm{Re}\, A_8$ directions.
If we wish to take care of the periodicity, we may use periodic
functional form in the neural network as in
Ref.~\cite{Alexandru:2017czx}.
In particular, violation of the periodicity
may become serious
when the
configurations are spread widely in the $\mathrm{Re} A_3$
and $\mathrm{Re} A_8$ variables.
For example, see Ref.\,\cite{Mori:2019tux} for the issue of the
periodicity.
As shown later, however, the present calculation
agrees well with
the mean-field approximation and configurations are well localized.
Thus, we do not introduce the periodic form of inputs at present.

It should be noted that the path optimization with the feedforward
neural network is unsupervised learning because we do not need teacher
data.
The setting of the feedforward
neural network in the path optimization method such as the optimizer are
the same with
Ref.~\cite{Kashiwa:2018vxr} and thus we skip the explanation here.

\section{Numerical results}
\label{Sec:NR}

In the actual numerical calculation, we have generated $80000$
configurations by using the hybrid Monte-Carlo method.
Then, the expectation values are estimated after a few times of
optimizations.
We employ the simple neural network which contains the input,
mono-hidden and output layers.
The number of unit in the hidden layer is set to $4 N_\mathrm{dof}=12$,
where $N_\mathrm{dof}$ is the number of dependent variables.
The expectation value of an operator (${\cal O}$) is obtained via the
phase reweighting as
\begin{align}
 \langle {\cal O} \rangle
 &= \frac{\displaystyle \int d^n t~{\cal O} e^{i\theta}|J(t) e^{-\Gamma(z(t))}|}
    {\displaystyle \int d^n t~e^{i\theta}|J(t) e^{-\Gamma(z(t))}| }
  = \frac{\langle e^{i\theta} {\cal O} \rangle_\mathrm{pq}}
         {\langle e^{i\theta} \rangle_\mathrm{pq}},
\end{align}
where $\langle \cdots \rangle_\mathrm{pq}$ means the phase quenched
average and
\begin{align}
 e^{i\theta} &= \frac{J(t) e^{-\Gamma(z(t))}}{|J(t) e^{-\Gamma(z(t))}|}.
\end{align}
The parameters in the NJL part are the same in Ref.\,\cite{Kashiwa:2018vxr} and
we newly introduce $G_v$ as $G_v=0.5 G$.

\begin{figure}[t]
 \centering
 \includegraphics[width=0.3\textwidth]{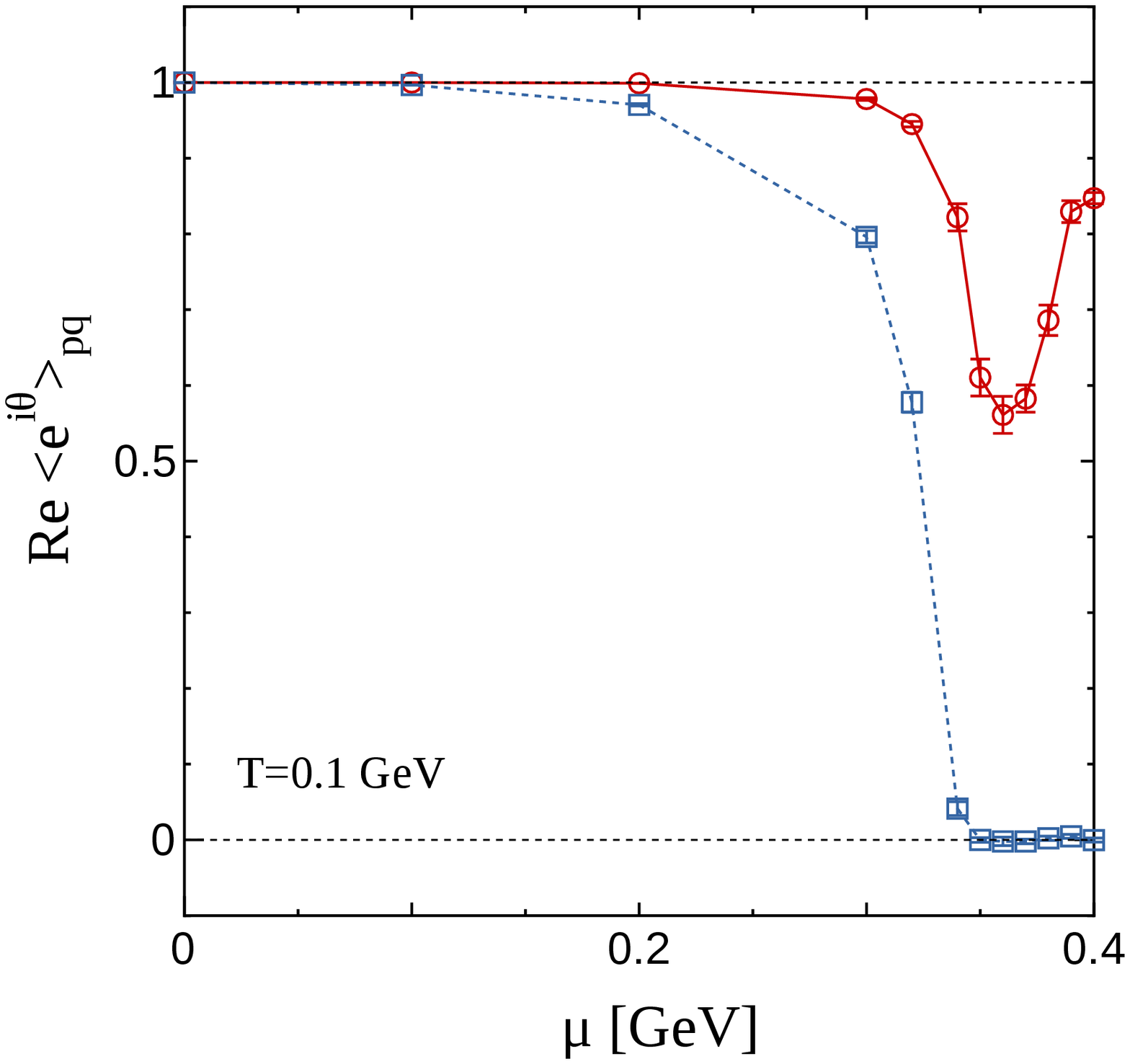}\\
 \vspace{5mm}
 \includegraphics[width=0.3\textwidth]{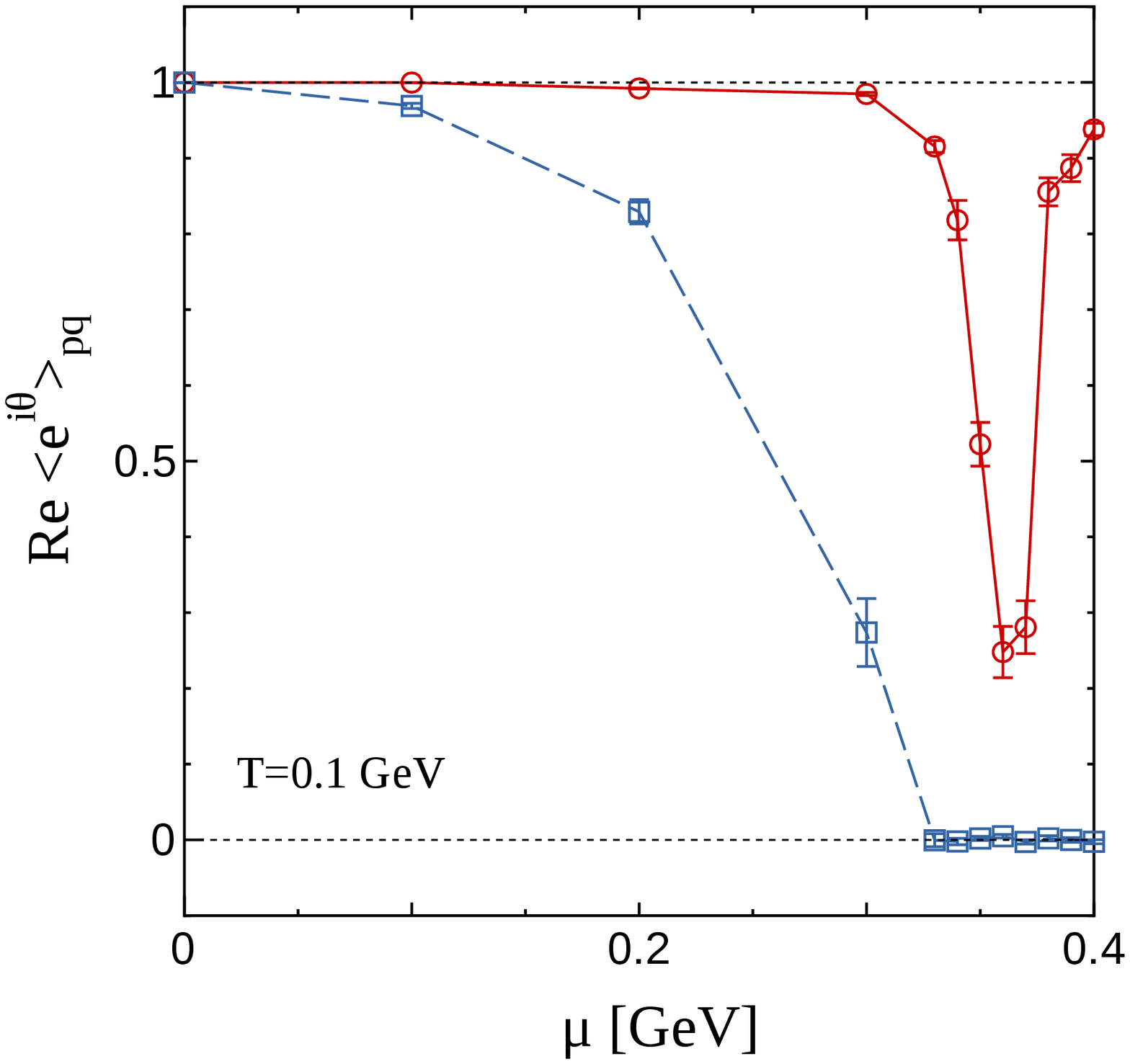}
 \caption{The $\mu$-dependence of
 $\mathrm{Re} \, \langle e^{i\theta} \rangle_\mathrm{pq}$ at $T=0.1$
 GeV.
 The top and bottom panels show results with $k=8$ and $64$,
 respectively.
 The circle and square symbols are results after and before the
 optimization, respectively.
 }
 \label{Fig:phase}
\end{figure}

Figure~\ref{Fig:phase} shows the average phase factor,
$\mathrm{Re}\,\langle e^{i\theta} \rangle_\mathrm{pq}$, at
$T=0.1$ GeV with $k= V T^3=8$ and $64$.
In some regions of $\mu$, the average phase factor becomes almost
$0$ before the optimization as shown by the dashed line in the figure.
By comparison,
we can successfully increase the average phase factor after the
optimization.
It suggests that there are no need to complexify
$\sigma$ and ${\vec \pi}$ auxiliary fields in the path optimization
method to investigate
the PNJL model with the repulsive vector-type interaction.
Also, this would be true in the Lefschetz-thimble method
and other complexified integral-path approaches.
Compared with the PNJL model without the repulsive vector-type
interaction~\cite{Kashiwa:2018vxr},
the average phase factor becomes worse because $\omega_4$
field additionally induces the sign problem at finite density.
Around $\mu=0.36$ GeV,
the optimization is neither sufficient nor automatic
in the case with $k=64$.
With naive initial conditions of dynamical variables, the average phase
factor stays to be very small.
Then, various initial conditions have been examined and we finally
obtain the optimized path with reasonably large average phase factor as
shown in Fig.~\ref{Fig:phase}.
This result indicates that the present neural network in the case with $k=64$
does not have enough performance of the approximation to overcome the
exponential suppression of the average phase factor.

 \begin{figure}[t]
 \centering
 \includegraphics[width=0.32\textwidth]{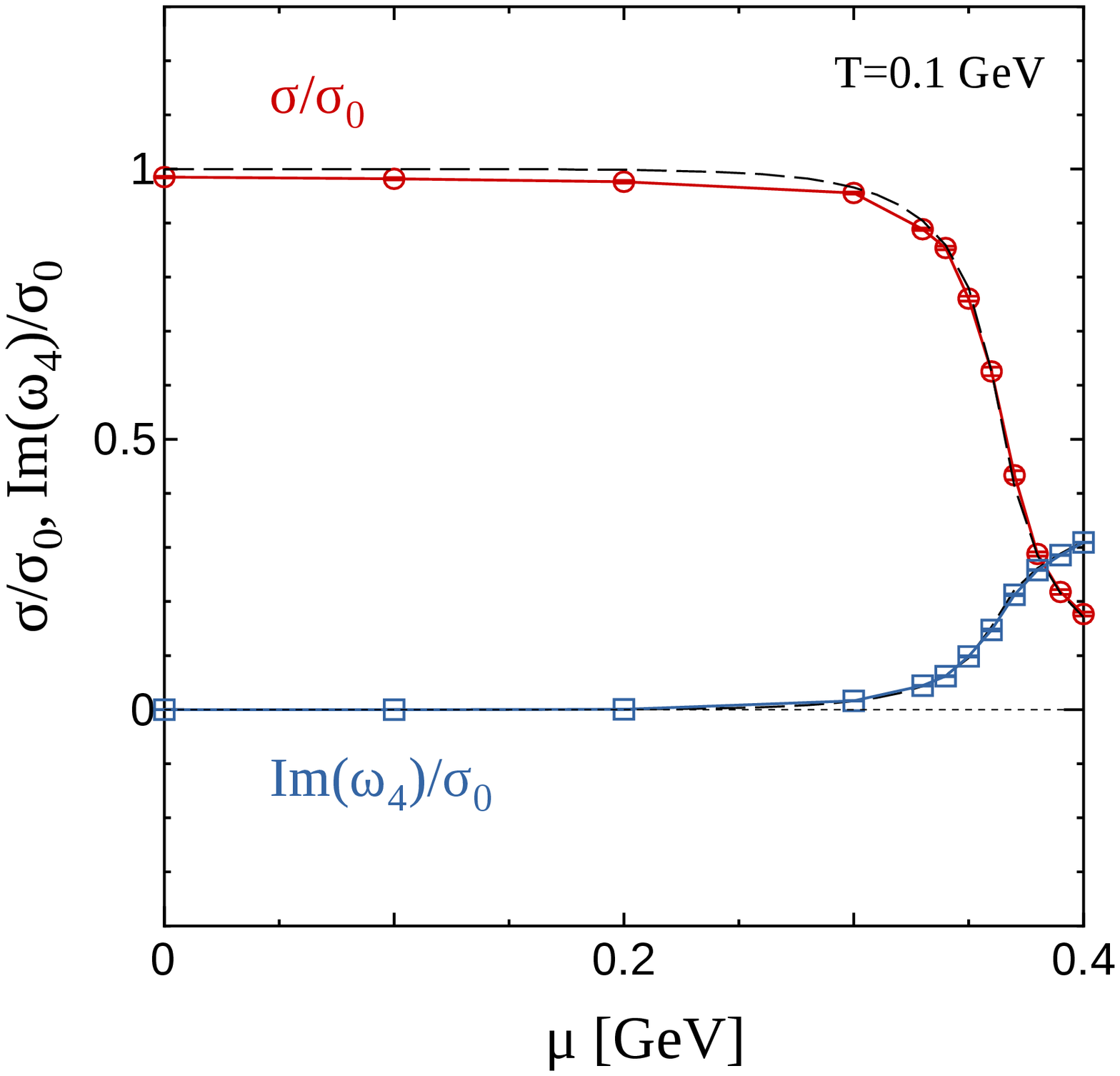}\\
 \vspace{5mm}
 \includegraphics[width=0.33\textwidth]{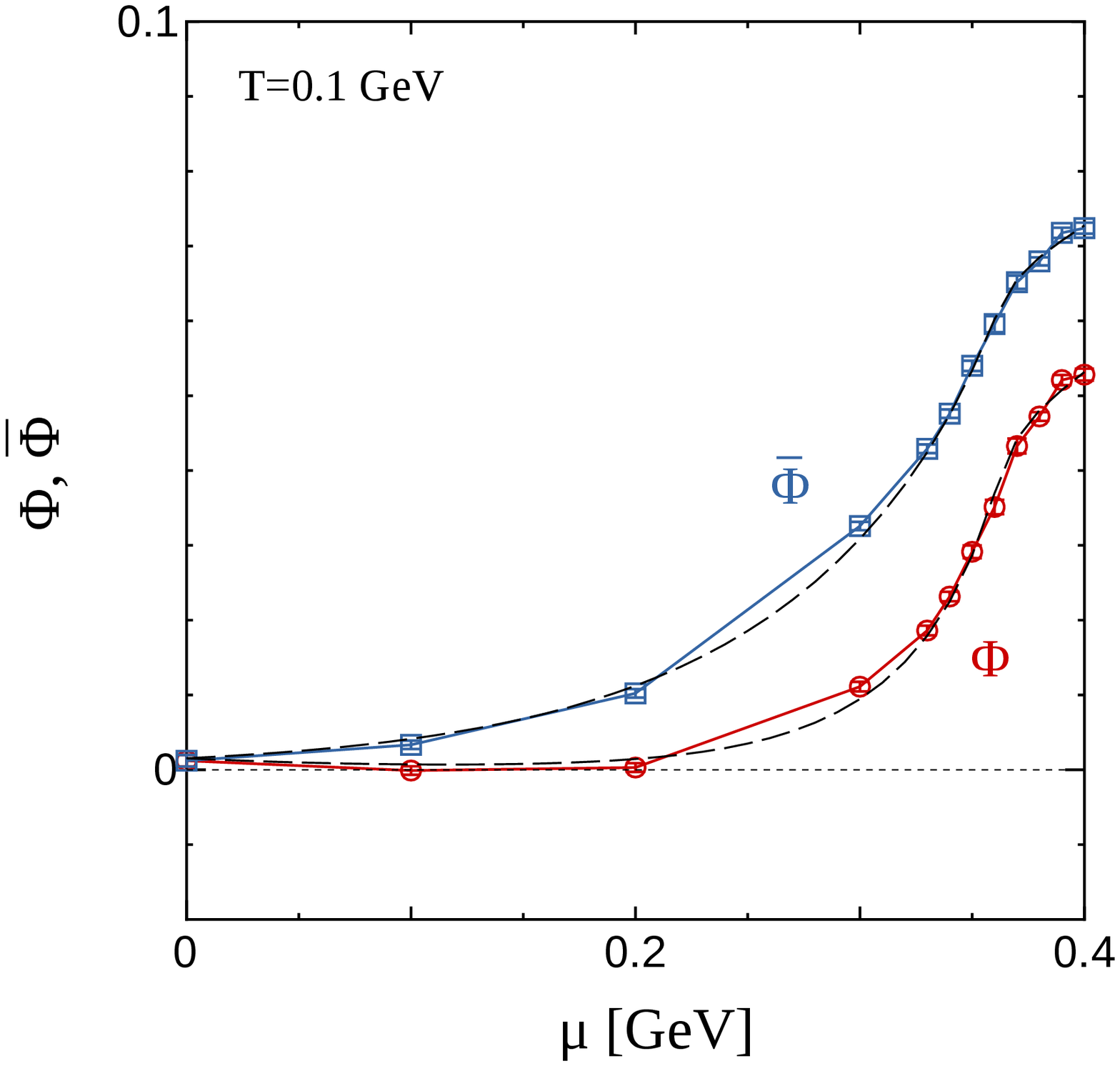}
 \caption{The top and bottom panels show the $\mu$-dependence of
 $\sigma$ and $\omega_4$, and $\Phi$ and ${\bar \Phi}$ at $T=0.1$
 GeV with $k=64$ were
 $\sigma$ and $\omega_4$ are normalized by
 $\sigma$ at $T=\mu=0$ ($\sigma_0 <0$) in the infinite volume limit.
 The thin dashed lines are the eye guide which are the mean-field results
  with the ${\cal CK}$
  symmetry ansatz in the fermion determinant and the un-Wick rotated
  calculation.
  }
 \label{Fig:T100}
\end{figure}
Figure~\ref{Fig:T100} shows the $\mu$-dependence of the order parameters
at $T=0.1$ GeV after the optimization.
We also show the mean-field results
based on the ${\cal C K}$ symmetry ansatz in the fermion
determinant~\cite{Nishimura:2014rxa,Nishimura:2014kla} where ${\cal C}$
and ${\cal K}$ are the charge and complex conjugations, respectively.
This ansatz can be justified
by using the Lefschetz thimble method~\cite{Tanizaki:2015pua}.
Under the ${\cal C K}$ symmetry condition,
we solve gap equations
\begin{align}
 \frac{\partial \Gamma}{\partial \sigma}=0,~~~~
 \frac{\partial \Gamma}{\partial \omega_0}=0,~~~~
 \frac{\partial \Gamma}{\partial \Phi}=0,~~~~
 \frac{\partial \Gamma}{\partial {\bar \Phi}}=0,~~~~
\end{align}
where we do not use the Wick rotation of the vector-type auxiliary field.
This treatment cannot be justified in the standard path integral formulation,
but practically it reproduces the correct result in the leading-order of
the large $N_c$ expansion because $\omega_0$ corresponds
to the quark number density in the mean-field approximation.
From the figure, we find that
the numerical errors are well controlled and the difference between $\Phi$
and ${\bar \Phi}$ at finite density, ${\bar \Phi} > \Phi$, is correctly
reproduced.
Compared with the results
without the vector-type interaction~\cite{Kashiwa:2018vxr},
the chiral condensate decreases more slowly.
This is reasonable, since the repulsive vector-type interaction
is known to weaken the chiral phase transition.
We find
that $|\mathrm{Im}\, \omega_4|$ strongly increases above $\mu=0.3$ GeV.
Since
the quark number density and $\omega_4$ are related with each other via
$\omega_0=\langle q^\dag q \rangle=i\omega_4$,
this sudden increase indicates
the absence of the Silver-Blaze problem at $T=0$;
the quark number density should start to increase at $\mu=M(\mu=0)$.
By comparison,
results at small $\mu$ are almost the same
as those without the vector-type interaction~\cite{Kashiwa:2018vxr},
since the quark number density and thus
the vector potential are small.

\begin{figure}[t]
 \centering
  \includegraphics[width=0.32\textwidth]{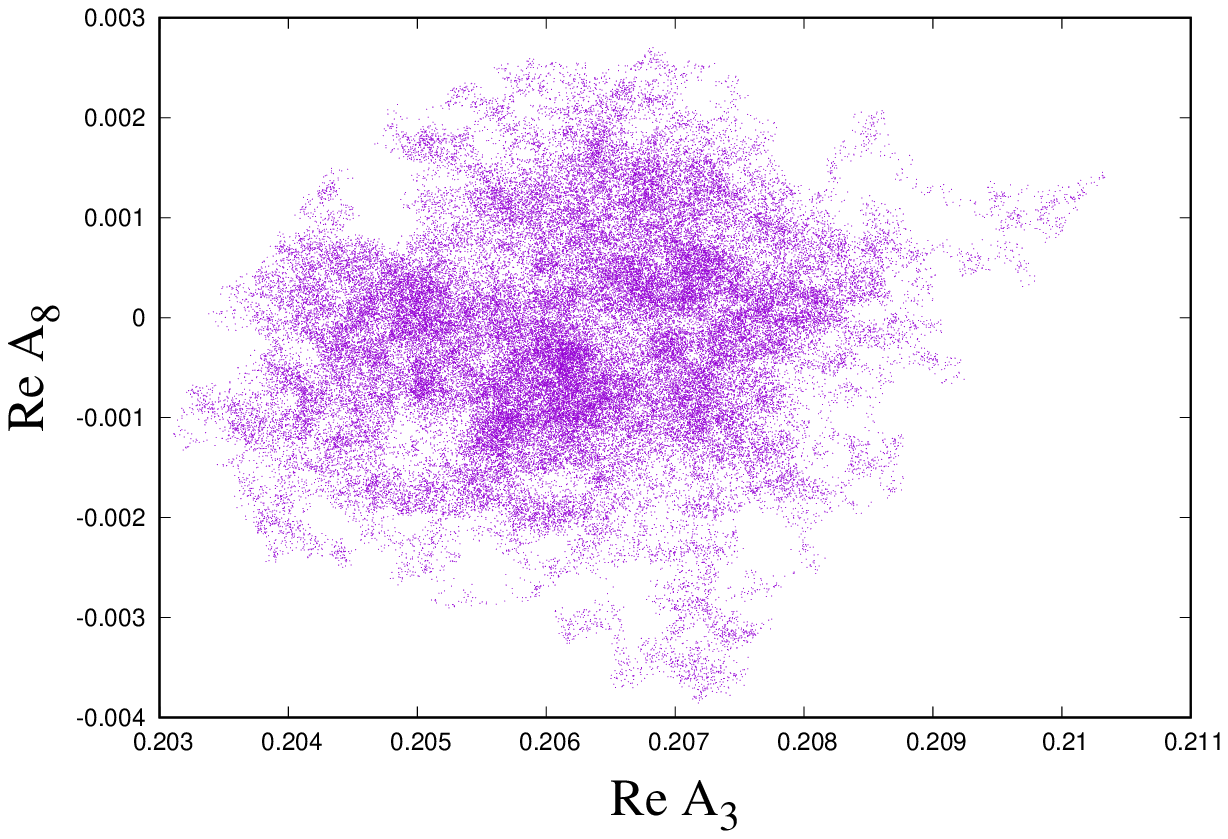}
  \includegraphics[width=0.32\textwidth]{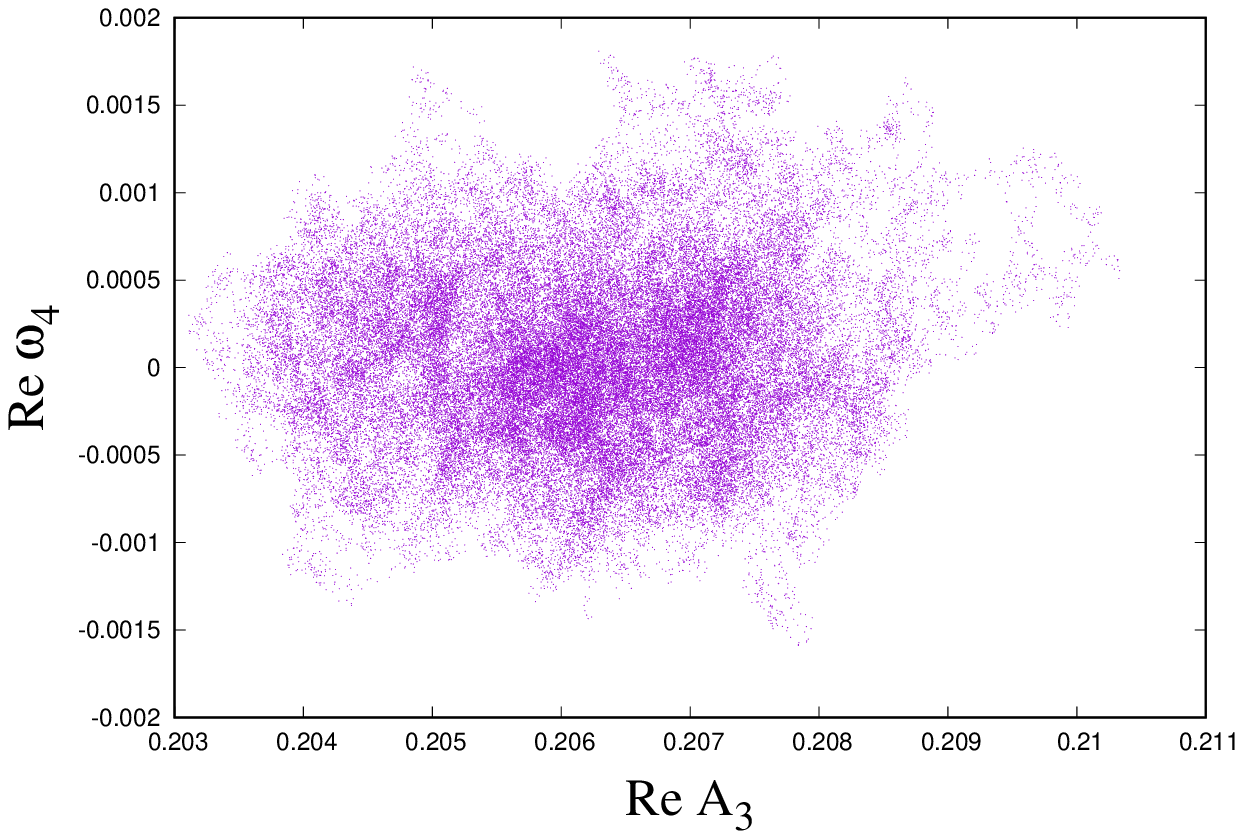}
  \includegraphics[width=0.32\textwidth]{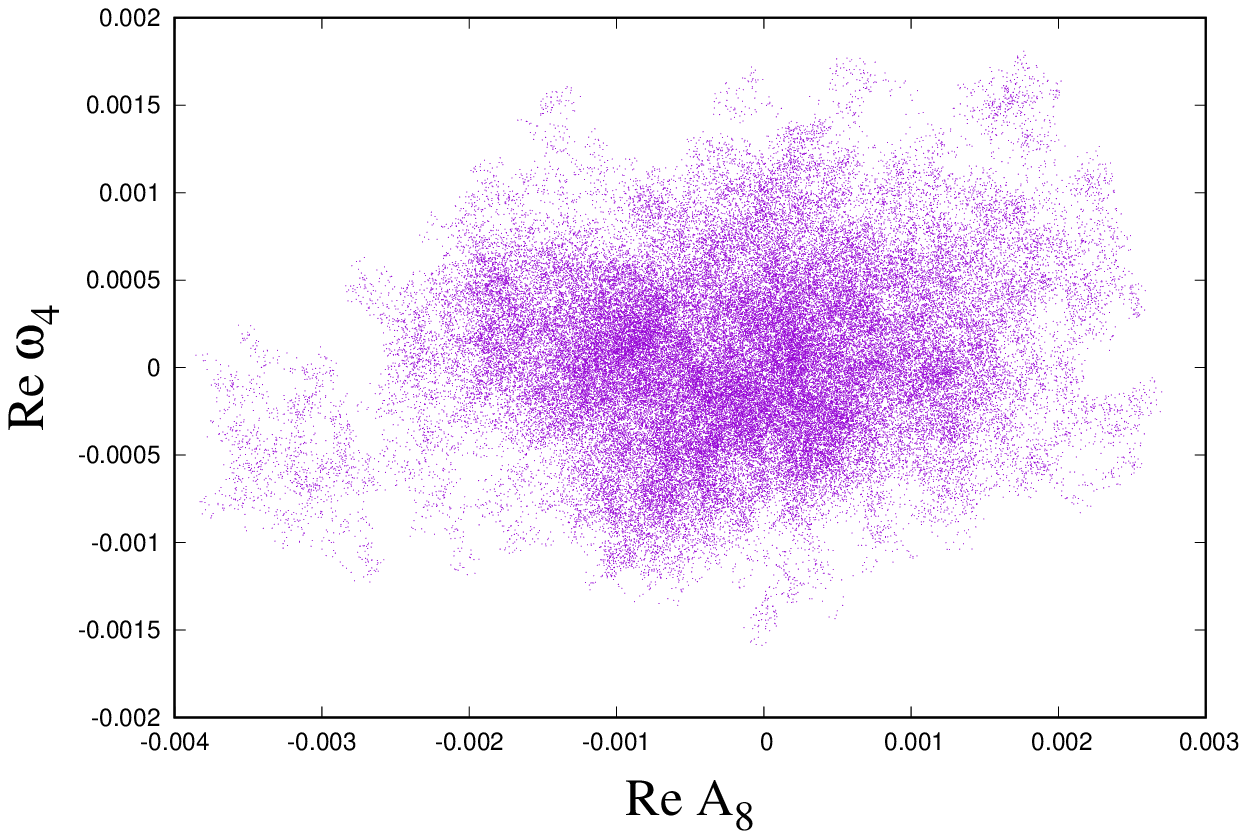}
  \caption{ The scatter plot at $T=0.1$ GeV and $\mu=0.3$ GeV
  on the
  $\mathrm{Re}\, A_3$-$\mathrm{Re}\, A_8$,
  $\mathrm{Re}\, A_3$-$\mathrm{Re}\, \omega_4$ and
  $\mathrm{Re}\, A_8$-$\mathrm{Re}\, \omega_4$ planes.
  Here we use $80000$ configurations.
  }
 \label{Fig:sp}
\end{figure}

It should be noted that the real part of $\omega_4$ is consistent with
zero within the error-bar and thus the consequence obtained in
the analyses in the Lefschetz thimble method
~\cite{Mori:2017zyl} is naturally understood
in the path optimization method.
We must consider Wick rotation of $\omega_0$,
then the model sign problem can be resolved by complexifying $\omega_4$.
The present results imply that the
$\omega_4$ field has almost only
the imaginary part, then the flow equation of the Lefschetz thimble
can stall at a small value of $|\mathrm{Re}\,\omega_4|$.
In Fig.~\ref{Fig:sp}, we
show the scatter plot
of the hybrid Monte-Carlo configurations
at $T=0.1$ GeV and $\mu=0.3$ GeV on the
$\mathrm{Re}\, A_3$-$\mathrm{Re}\, A_8$,
$\mathrm{Re}\, A_3$-$\mathrm{Re}\, \omega_4$ and
$\mathrm{Re}\, A_8$-$\mathrm{Re}\, \omega_4$ planes.
We can see the localized configurations around
$\mathrm{Re}\, A_8 =0$ and $\mathrm{Re}\, A_3 \neq 0$.
Thus, this study implies
that the standard PNJL model computation
with the repulsive vector-type interaction under the ${\cal CK}$ symmetry
ansatz and the un-Wick rotated $\omega_0$ in the mean-field
approximation is
systematically and numerically justified via the path optimization
method.
The scatter plot also supports the direct complexification
of $A_3$ and $A_8$ without taking care of the periodicity.
The Monte-Carlo configurations are not spread but localized,
then it is not necessary to take account of the periodic boundary condition.

\section{Summary}
\label{Sec:Summary}

In this study, we have applied the path optimization method to the QCD
effective model with the Polyakov loop and the repulsive vector-type
interaction.
The feedforward neural-network with the mono-hidden layer is employed
to describe the good integral
path in the complexified space of integral variables.
The temporal components of the gluon field and the vector-type auxiliary
field are complexified and then the path is optimized via the path
optimization method.

By optimizing the path (manifold),
we can successfully improve the average phase factor,
and calculated results of observables
show reasonable behavior and have small error bars.
It is not easy to optimize the integral path in the rapidly
changing region of the order-parameters,
but we can finally improve the average phase factor
by examining various initial conditions of dynamical variables.

After a few optimization steps, we can well reproduce the
mean-field results at large volume as we expect.
Since we use the homogeneous ansatz of the integral valuables,
our numerical simulation should give the mean-field result
in the large volume limit.
The imaginary part of the vector-type auxiliary field starts to rapidly
increase in strength above $\mu=0.3$ GeV at $T=0.1$ GeV.
This indicates the absence of the Silver-Blaze problem at $T=0$ and
thus the path
optimization method can pick up correct properties of the theory.

In the standard mean-field approximation, we do not perform the Wick
rotation of the vector auxiliary field ($\omega_0$).
While such a treatment cannot be justified
within the standard path integral formulation,
it can be justified by employing the
complexified theory such as the Lefschetz thimble method and
the path optimization method.
In this article, we have demonstrated that the path optimization method
correctly resolve the model sign problem and then the $\omega_4$ field takes
almost the pure imaginary value which is required from
the fact that the grand-canonical partition function is real.
This study provides the correct numerical treatment
of the repulsive vector-type interaction in the QCD effective model with the
Polyakov loop.

Finally, we comment on the problem of the numerical cost.
The degree of freedom is enlarged in the present calculation
compared with the case without vector-type
interaction~\cite{Kashiwa:2018vxr}, the improvement of the
average phase factor becomes slow, and thus we need more optimization
steps (epochs) and/or some other extensions.
One of the possible extensions to circumvent such optimization problem is
introducing the deep neural-network and it is our future work.
Also, the sign problem becomes exponentially severe with increasing system size.
Then, it is important to know that the improvement of the average phase
factor via the path optimization method can overcome the exponential
suppression.
Therefore, we need further investigation of the competition
in the average phase factor between the
suppression from the system size and the improvement from the path
optimization.
In particular, this problem becomes serious when we consider the lattice
calculation.
One promising approach is the
reduction of the Jacobian computation cost;
the diagonal ansatz of the Jacobian
~\cite{Alexandru:2018fqp} and the nearest-neighbor
lattice-cites ansatz~\cite{Bursa:2018ykf} are promising examples.
It will be reported elsewhere.

\begin{acknowledgments}
This work is supported in part by the Grants-in-Aid for Scientific Research
 from JSPS (Nos. 15H03663, 16K05350, 18J21251, 18K03618, 19H01898),
 and by the Yukawa International Program for Quark-hadron Sciences (YIPQS).
\end{acknowledgments}

\bibliography{ref.bib}

\end{document}